# Large Magnetoresistance and Weak Antilocalization in $V_{1-\delta}Sb_2$ Single Crystal

Yong Zhang[1,2], Xinliang Huang[2], Wenshuai Gao[1]*, Xiangde Zhu[2]*, Li Pi[2]

[1]Institutes of Physical Science and Information Technology, Anhui University, Hefei 230601, China

[2]Anhui Province Key Laboratory of Condensed Matter Physics at Extreme Conditions, High Magnetic Field Laboratory of the Chinese Academy of Sciences, Hefei 230031, Anhui, China

*gwsh@ahu.edu.cn

*xdzhu@hmfl.ac.cn

**Abstract: The binary pnictide semimetals have attracted considerable attention due to their fantastic physical properties that include topological effects, negative magnetoresistance, Weyl fermions and large non-saturation magnetoresistance. In this paper, we have successfully grown the high-quality $V_{1-\delta}Sb_2$ single crystals by Sb flux method and investigated their electronic transport properties. A large positive magnetoresistance that reaches 477% under a magnetic field of 12 T at T = 1.8 K was observed. Notably, the magnetoresistance showed a cusp-like feature at the low magnetic fields and such feature weakened gradually as the temperature increased, which indicated the presence of weak antilocalization effect (WAL). The angle-dependent magnetoconductance and the ultra-large prefactor α extracted from the Hikami-Larkin-Nagaoka equation revealed that the WAL effect is a 3D bulk effect originated from the three-dimensional bulk spin-orbital coupling.**

# 1. Introduction

Topological insulators have attracted significant attention as they possess helical surface states protected by time reversal symmetry.[1-6] Due to the topological origin and Dirac fermion nature of the topological surface states, the topological insulators are expected to perform excellently in transport behaviors.[7, 8] After the discovery of fascinating topological insulators, the binary pnictide semimetals $XPm_2$ (X =V, Nb, Ta; Pm =As, Sb)[9] (listed in Table I) have also became a focus of attention in the field of condensed matter. The electronic structures of such materials are protected by an interplay of symmetry and topology. Electronic transport shows many novel quantum phenomena such as large non-saturation magnetoresistance (*MR*), [10-16] ultrahigh carrier mobility,[17] negative *MR*.[18-20] These multifunctional characteristics not only provide us a promising platform for fundamental physical research but also open up a new route for exploring the potential spintronic applications which is one of the most important reasons why these materials have stimulated unprecedented research.[21]

$V_{1-\delta}Sb_2$ has long been known as thermoelectric materials.[22] It has been recognized that this type of material is one of the candidates for diffusion barriers or electrodes in thermoelectric devices based on skutterudites.[6] So far, there is very little research on $V_{1-\delta}Sb_2$. In this work, we successfully grow the large size and high-quality single crystals of $V_{1-}$

$_\delta$Sb$_2$ single crystals. We systematically characterize its crystal structure, electrical transport, magnetism properties, specific heat, and calculated band structure.

## 2. Experimental details

The high-quality V$_{1-\delta}$Sb$_2$ single crystals were synthesized by self-flux method[23]. V dendritic crystals (4N) and Sb (5N) pieces with V : Sb molar ratio of 1 : 30 were mixed and held in an alumina crucible, and sealed in an evacuated quartz tube. The quartz tube was heated to 1100 ˚C in 10 hours. After holding at this temperature for 20 hours, the tube was slowly cooled to 700 ˚C at a rate of 3 ˚C/h. The tube was then removed from the furnace and inverted into a centrifuge to remove the excess molten antimony. The resulting metallic luster single crystals with several millimeters size were collected from the crucible. The typical dimensions of the V$_{1-\delta}$Sb$_2$ single crystals are 1 mm×1mm×0.5mm and show no signs of instability or degradation when exposed to air. The compositions and elemental stoichiometry of the crystals were determined by energy-dispersive spectrometer (EDS) on Oxford X-MAX electron spectrometer equipped on the focused SEM/FIB dual beam system (Helios Nanolab600i, FEI Inc.).

A single crystal was selected and data were collected on a Bruker D8 Quest diffractometer with a detector using a Cu K$\alpha$ radiation source ($\lambda$ =

1.54184 Å) from a rotating anode in a range of 10-90°. Magnetization data were obtained from a Quantum Design MPMS Superconducting Quantum Interference Device (SQUID) magnetometer with a 7 T superconducting magnet.

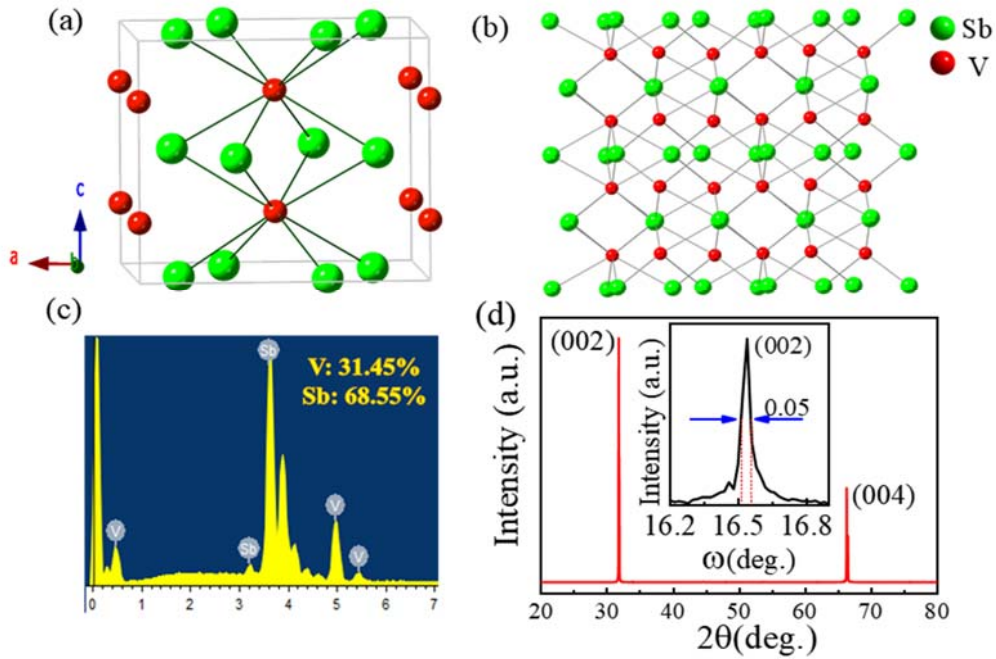

**Fig. 1.** (a) and (b) Crystal structure of the $V_{1-\delta}Sb_2$ single crystal. The red sphere represents vanadium atoms, the green sphere represents antimony atoms. (c) The topical Energy Dispersive Spectrometer (EDS) for single crystal $V_{1-\delta}Sb_2$. (d) X-ray diffraction pattern of $V_{1-\delta}Sb_2$ single crystal with the corresponding Miller indices (00L) in parentheses. The inset shows a corresponding rock curve.

The magnetization measurements of $V_{1-\delta}Sb_2$ single crystal were performed with zero-field cooling (ZFC) process on MPMS. Electrode was made by spot welding using 25 μm gold wire, and electrical transport measurements were carried out by the four-probe method upon a physical property measurement system (PPMS) Quantum Design at a temperature range from 1.8 K to 300 K. During the measurements, a model 372 (AC) resistance bridge and temperature controller were used for the measurement of resistance with applied current of 3.16mA. The low-temperature specific heat measurements were carried out upon a 16 T physical properties measurement system (Quantum Design, PPMS-16T) with a dilution refrigerator using the standard thermal relaxation method.

The preliminary first-principles band structure calculations were performed based on the density functional theory (DFT), and use the projector augmented wave method as implemented in the Vienna ab initio simulations package (VASP).[24] The generalized gradient approximation with the Perdew-Burke-Ernzerhof (PBE)[25] realization was adopted for the exchange-correlation functional. The cutoff energy was set as 400 eV. The energy and force convergence criteria were set to be $10^6$ eV and -0.05 eV/Å, respectively.

## 3. Results and discussion

According to the previous report,[26] $CuAl_2$ type structure $V_{1-\delta}Sb_2$ crystallizes in the tetragonal *I4/mcm* space group (NO. #140). Figure 1(a) and (b) show the crystal structure of $V_{1-\delta}Sb_2$,[26] which can be regarded as alternating stacking layers of V and Sb atoms by ABAB sequence along the *c*-axis. Each V atom is octahedrally coordinated by Sb atoms with a bond distance of 2.484(8) Å.[27, 28]

Figure 1(c) shows the EDS spectrum for $V_{1-\delta}Sb_2$, the measurements were performed at different positions on the polishing surface of the crystal. The average chemical composition ratio of V : Sb was established to be 31.45%:68.55%, which confirms the existence of V atoms vacancies in the $V_{1-\delta}Sb_2$ single crystal with $\delta \approx 0.08$. This value is approximately in accordance with previous studies($V_{0.9}Sb_2$[29] and $V_{0.97}Sb_2$[6, 22]). This is the reason why $V_{1-\delta}Sb_2$ crystal perform tetragonal structure, rather than monoclinic structure as other pnictides $XPm_2$ (as shown in Tab. I).

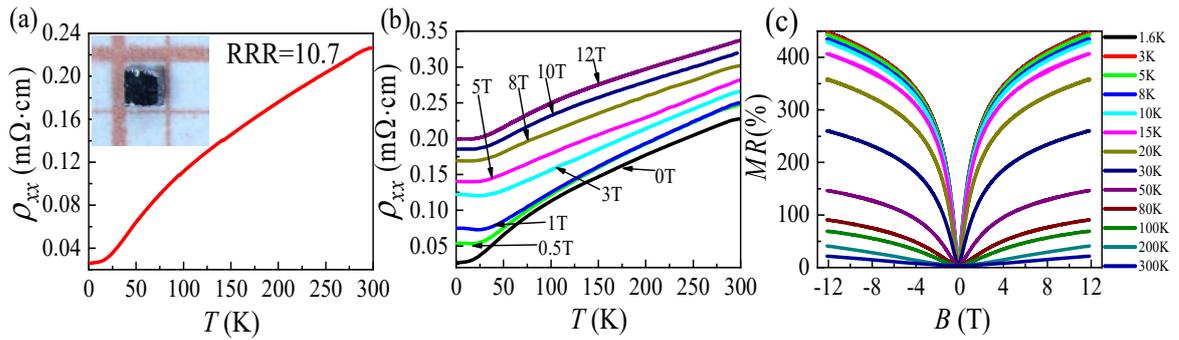

**Fig. 2.** (a) The temperature dependence of longitudinal $\rho_{xx}$ from 1.8 K to 300 K at the zero magnetic field. The value of *RRR* is 10.7. Upper inset shows a photograph of $V_{1-\delta}Sb_2$ single crystal on millimeter-grid paper. (b) The $\rho_{xx}$ at different magnetic field. (c) The normalized *MR* at temperature ranging from 1.8 K to 300 K.

Figure 1(d) shows the single crystal XRD pattern for the $V_{1-\delta}Sb_2$ at room temperature. Two sharp diffraction peaks that represent (*00l*) plane can be observed. The inset shows the rock curve of the *(002)* diffraction peak, the full-width-at-half-maximum (FWHM) is estimated to be $\Delta\theta \approx 0.05°$. Such a narrow FWHM indicates the high quality of the single crystal without twin crystal.

Figure 2(a) exhibits the temperature dependence of longitudinal resistivity $\rho_{xx}$ for V$_{1-\delta}$Sb$_2$ single crystal in the temperature range of 300 K-1.8 K without magnetic field. The inset displays the typical view of a V$_{1-\delta}$Sb$_2$ single crystal placed on a millimeter grid with the preferential orientation along the [*001*] zone axis. The residual resistance ratio (*RRR*),$RRR = \rho_{xx}(300K)/\rho_{xx}(2K) = 10.8$, is relatively lower than that of VAs$_2$(12),[17] NbAs$_2$(150)[30] and NbSb$_2$(~200).[31] Such value is still larger than the reported value of ~3.0 in Ref [6]. Such low value of *RRR* for V$_{1-\delta}$Sb$_2$ should origin from the intrinsic V vacancy, which provides additional scattering centers.

When the temperature decreases from 300 K to 1.8 K, the $\rho_{xx}$ exhibits metallic behavior. Figure 2(b) plots the temperature dependence of resistivity under different magnetic field. The electric current is parallel to the *b*-axis, and the magnetic field is perpendicular to the *ab* plane. When a magnetic field is applied to the V$_{1-\delta}$Sb$_2$ single crystal, the $\rho_{xx}$ curves are enhanced obviously with increased magnetic. However, the V$_{1-\delta}$Sb$_2$ still holds metallic behavior even at high field of 12T without any sign of metal-insulator transition, which is quite different from the observation of insulator-like behavior in other binary pnictide semimetals. [13, 14, 31-33]

Figure 2(c) shows the normalized *MR* curves measured at different temperatures and the magnetic field is applied perpendicular to the (*001*) plane. The *MR* is defined as $MR=\frac{\rho_{xx}(H)-\rho_{xx}(0)}{\rho_{xx}(0)} \times 100\%$. Here $\rho_{xx}(H)$ and

$\rho_{xx}(0)$ are longitudinal resistivities with and without magnetic field, respectively. The normalized *MR* curves are symmetrized via *MR(B)*=[*MR(B)* + *MR(-B)*]/2 to remove the contribution of Hall resistance. At T=1.8 K, $V_{1-\delta}Sb_2$ single crystal exhibits a large *MR* of 447% at B=12 T. With the increase of temperature, *MR* is suppressed significantly but still remain about 18% at 300 K. Remarkably, we observed that the *MR* increases sharply in the low field region under 20 K, and a large *MR* of 180% is obtained in a magnetic field of 1 T at 1.8 K. The *MR* in $V_{1-\delta}Sb_2$ single crystal is much larger than many other nonmagnetic compounds, such as ScPdBi,[34] LuPtSb[35] and $Bi_2Te_3$.[36] This distinct *MR* dip near zero magnetic field can be associated with the weak antilocalization (WAL) effect. The WAL effect is generally expected to originate from the topologically protected surface state,[37-39] or strong spin-orbit coupling.[40] At zero magnetic field, backscattering gains the minimum due to the time-reversal symmetry. As the magnetic field increases, the time-reversal symmetry is broken and the backscattering increases, which leads to a sharp increase of *MR*. It has been widely observed in topological insulators (Bi-Se-Te[41-43]) and topological superconductors (Al-InAs[44]and LuPdBi[45]). Furthermore, such phenomenon has also been reported in systems with strong spin-orbit couplings, such as LuPtSb[35] and YPdBi.[46]

In order to deeply understand the WAL effect in $V_{1-\delta}Sb_2$ single crystal, we fit the corresponding magnetoconductivity $\Delta\sigma_{xx}(B) = \sigma_{xx}(B) -$

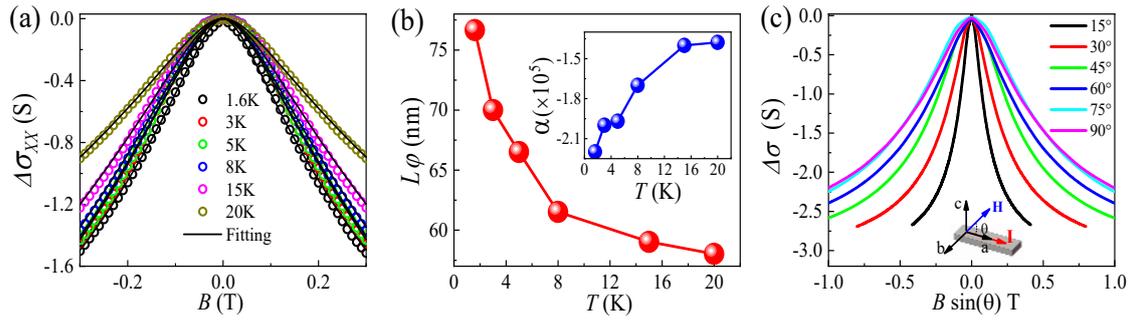

**Fig. 3.** (a) The normalized magnetoconductance ($\Delta\sigma_{xx}$) at different temperature (color hollow circle) and the black line is corresponding fitting curve. (b) Temperature dependence of the fitting parameter $L\varphi$ phase coherence length deduced from the WAL fit. The inset shows the temperature dependence of $\alpha$. (c) The $\Delta\sigma_{xx}$ plotted in the perpendicular magnetic field component of the

$\sigma_{xx}(0)$ using the Hikami-Larkin-Nagaoka (HLN) formula[47]

$$\Delta\sigma_{xx}(B) = \frac{\alpha e^2}{2\pi^2 \hbar}\left[\psi\left(\frac{1}{2}+\frac{\hbar}{4eBL_\varphi}\right)-\ln\left(\frac{\hbar}{4eBL_\varphi}\right)\right] \dots\dots\dots\dots\dots (1)$$

where $e$, $\hbar$, are the electron charge and reduced plank constant; $\psi$ is digamma function; $\alpha$ and $L_\varphi$ are the prefactor and phase coherence length, respectively. The prefactor $\alpha$ denotes the number of conduction channels, sswhich is experimentally found to be -0.5 for each conductive channel in two-dimensional (2D) topological insulator. Figure 3(a) displays the low-field magnetoconductivity $\Delta\sigma_{xx}(B)$ from -0.3 T to 0.3 T at temperatures ranging from 1.8 K to 20 K, and all the curves can be well fitted by the HLN model. The obtained phase coherence length $L\varphi$ as a function of temperature are shown in Figure 3(b), and the inset displays the temperature dependence of parameters α. Remarkably, the value of α with an order of $10^5$ is larger than the theoretical 2D electron system, hinting that the WAL effect is mainly originate from the strong spin-orbit coupling of the three-dimensional (3D) bulk state in the $V_{1-\delta}Sb_2$ single crystal. In addition, we measure the angle-dependent $\Delta\sigma_{xx}(B)$ at 1.8 K under low magnetic fields from -1 T to 1 T, as shown in Figure 3(c). In the 2D

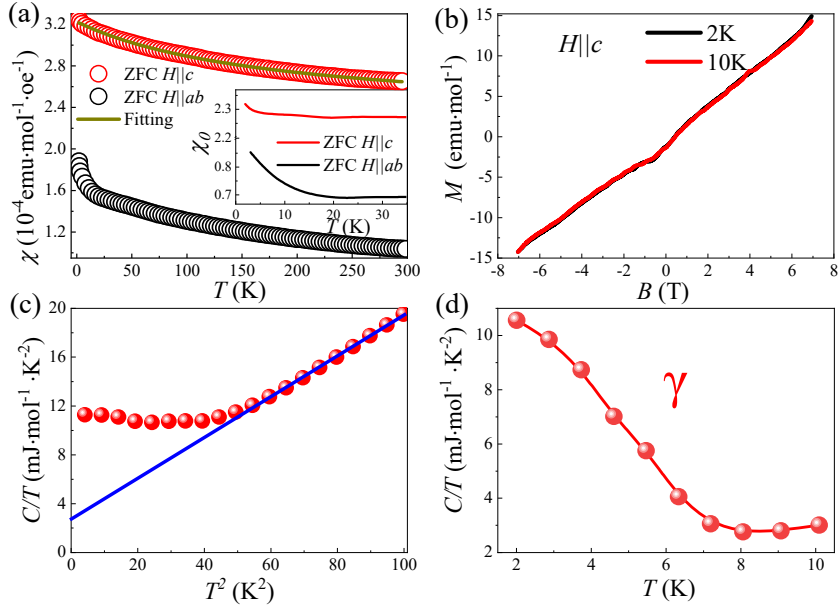

**Fig. 4.** (a) The magnetic susceptibility of $V_{1-\delta}Sb_2$ single crystal with under zero field cooling (ZFC) for field parallel to *c*-axis and perpendicular to *ab* plane. The brown line is fitting with the Curie-Weiss law and the inset shows is a temperature dependent paramagnetic susceptibility gained by magnetic susceptibility subtracted the Curie-Weiss part. (b) Field dependent magnetization at 2K, 10K, respectively. (c) $C(T)/T$ vs. $T^2$ of $V_{1-\delta}Sb_2$ single crystal from 2 K to 10 K. (d) The free electron part coefficient $\gamma$ ($C/T$-$\beta T^2$) versus temperature *T*, where $\beta$ is determined from the linear fitting (blue solid line).

topological insulator, the $\Delta\sigma_{xx}$ vs $B\sin\theta$ curves should converge to one curve. The $\Delta\sigma_{xx}$ vs $B\sin\theta$ curves in $V_{1-\delta}Sb_2$ are separate from each other, which further confirms the 3D feature of the WAL effect. All the results reveal that the WAL effect in $V_{1-\delta}Sb_2$ is resulted from strong spin-orbit couplings destructive quantum interference between time-reversed loops formed by scattering trajectories, leading to the conductivity enhanced with

decreasing temperature in the 3D bulk transport.[39]

In addition, we measured the anisotropic magnetic properties of the V$_{1-\delta}$Sb$_2$ single crystal. Figure 4(a) shows the magnetic susceptibility ($\chi$)-T curve with applied magnetic field parallel and perpendicular to (*001*) plane, respectively. Obviously, the temperature dependent paramagnetic susceptibility curves show typical paramagnetic behavior with nonnegligible anisotropic, and the $\chi_c$ is relatively larger than $\chi_{ab}$. Both curves can be well fitted with the Curie-Weiss law $\chi = \chi_0+C/(T-\theta)$,[48] where $\chi_0$ is the temperature dependent paramagnetic susceptibility from the electron contribution near the Fermi level, $C$ is Curie constant, $\theta$ is the Weiss temperature. For both curves, the determined $C$ and $\theta$ are almost identical with the values of about 0.0174(9) emu mol$^{-1}$ Oe$^{-1}$ K$^{-1}$ and -183.7(3) K, respectively. The effective momentum determined to be 0.3741 $\mu_B$/V atom. The negative $\theta$ value indicates the existence of antiferromagnetic coupling between V atoms or vacancy. By subtracting the Curie-Weiss part, the $\chi_0$ part can be obtained and is shown in the inset of Fig. 4(a). Apparently, $\chi_0$ for H||c remains in whole temperature range down to 4 K with small curie tail, while $\chi_0$ for H||ab starts to increase suddenly with decreasing temperature below 15K. Such a sudden increase of $\chi_0$ for H||ab may not come from magnetic order, since the antiferromagnetic coupling. As shown in Fig. 4(b), the magnetic hysteresis loop at low temperature exhibits the typical paramagnetic behaviors for

*H*//c, which confirmed that an abrupt increase of the $\chi_0$ part is not induced by the magnetic order. In classic condensed matter physics, the $\chi_0$ is proportional to the density of fermi level with small magnetic field *H* with relation of

$$\chi = \frac{N \mu_0 \mu_B^2}{k_B T_F} \text{[49]} \quad \quad \quad \quad \quad \quad \quad \quad \quad \quad (2)$$

where *N* represents the density of states at the fermi level, $\mu_0$ is the magnetic permeability of free space, $\mu_B$ is the Bohr magneton, $k_B$ is the Boltzmann's constant, and *T* is the temperature, and $T_F$ is the fermi temperature. Therefore, anisotropic increase of $\chi_0$ suggests that the density of states *N* increase abruptly at low temperature. Since the electronic heat capacity part $\gamma$ is proportional to $N^{1/3}$ and effective mass $m^*$ at low temperature with equation

$$\gamma = \frac{m^*}{m} * \frac{4\pi^3 m k_B^2}{3h^2} \left(\frac{3NV^2}{\pi}\right)^{\frac{1}{3}} \text{[50]} \quad \quad \quad \quad \quad \quad (3)$$

Where *m* is the mass of free electrons, *h* is Planck constant, *V* is the volume per mole. Theoretically, an abrupt increase of *N* at low temperature is expected to yield an abrupt increase of $\gamma$ in specific heat.

Heat capacity was measured between 2 K to 10 K and no phase transition was observed. The total heat capacity of quantum material at low temperature can be described as the sum of $C = C_e + C_p = \gamma T + \beta T^3$, where $C_p$ is the phonon contribution, $\gamma$ and $\beta$ are the free electron part coefficient and phonon part coefficient.[51] To obtain the $\gamma$ part, $C/T$ versus $T^2$ curve is plotted in the Figure 4(c). By linear fitting of the curve, phonon part coefficient $\beta$=0.16529 mJ mol$^{-1}$ K$^{-4}$ can be obtained. After subtracting the phonon part, the temperature dependent of $\gamma$ can be obtained, as displayed in the Fig. 4(d). Obviously, the $\gamma$ free electron part suddenly increase under the 7 K, which confirm that the abrupt increase of $\gamma$ induced by the abrupt increase of $N$ at low temperature.

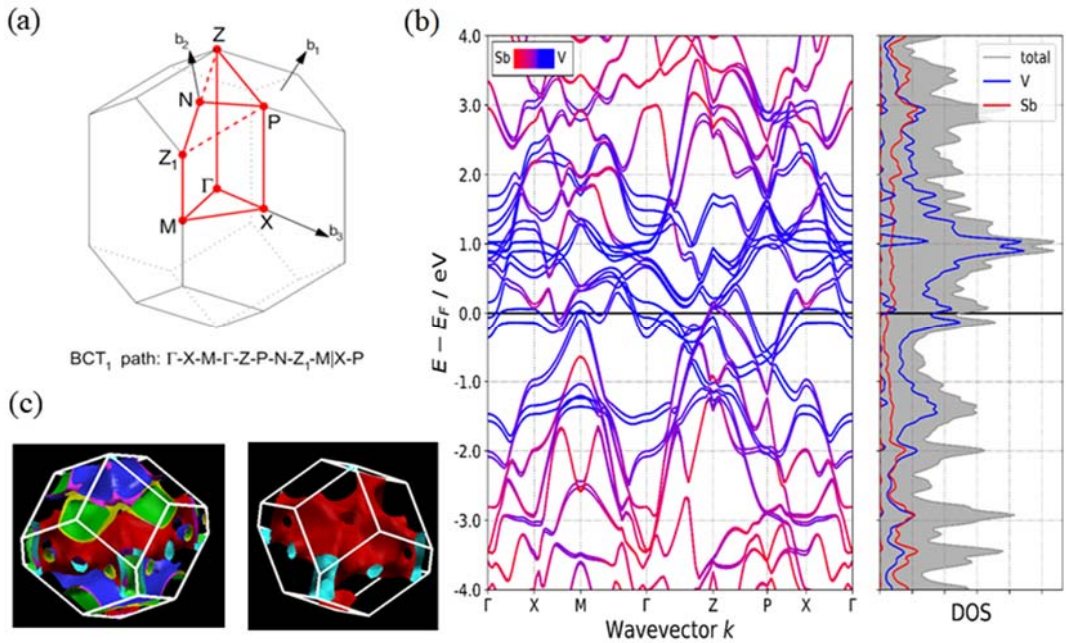

**Fig. 5.** (a) The Brillouin zones. (b) Band structures calculated with considering spin-orbit coupling (SOC) and the total density of states (DOS) for VSb$_2$. (c) The Fermi surfaces of VSb$_2$ crystals from first-principle calculation. Left is total fermi surface for VSb$_2$, right is fermi surface with along Γ-X.

We first calculate the band structures with SOC. The obtained results are plotted in Figure 5(b), Then, Fermi Surface was calculated by the VASPKIT [24]. The software of XcrySDen was used to plot the Fermi Surface in Figure 5(c). Figure 5(a) shows the first Brillouin zone in momentum space for the primitive unit cell. The metallic character can be shown by the high density of states (DOS) at the Fermi level and the presence of electron and hole-bands. The DOS of $VSb_2$ show that the region close to Fermi level is dominated by fully occupied V-3d states with a small contribution of Sb-3p states and that conduction bands are dominated strongly by this hybridization.[26] The left of Figure 5(c) reveals the total fermi surface include the six bands crossing fermi level. The right picture shows a flat band, that cross the fermi surface along Γ-X point, and the X point is in the center of the hexagonal plane of Brillouin zone. When the presence of V atoms deficiency in the $VSb_2$ single crystal, the fermi surface is likely cross the flat band and produce the large $N$. This is the reason why the magnetic susceptibility and the $C_e$ increase at low temperature.

## 4. Conclusions

High quality $V_{1-\delta}Sb_2$ single crystals were grown via flux method. Due to $VSb_2$ possesses a narrow homogeneity range, the vanadium compound possesses a V deficiency and is better described as $V_{1-\delta}Sb_2$ ($\delta \approx 0.8$). It is the possible reason why the property of $V_{1-\delta}Sb_2$ is different from other binary pnictide semimetals. We have calculated the electronic structure with SOC and measured the longitudinal resistivity, specific heat, and magnetic susceptibility of $V_{1-\delta}Sb_2$. Transport experiments demonstrated that $V_{1-\delta}Sb_2$ performs metallic behavior with positive large *MR* reached 447%. More intriguingly, we observed an obvious WAL effect from MR below 20 K in the low-field region. The extremely large $\alpha$ and the angle dependence of $\Delta\sigma_{xx}$ suggest that the WAL effect in $V_{1-\delta}Sb_2$ single-crystal originates from the contribution of strong 3D bulk spin-orbital coupling. In addition, we observed the abruptly increase of the $\chi_0$ part and $C_e$ part, which is according with the characteristics for the flat band. Such band structure indicates $V_{1-\delta}Sb_2$ is likely to display fantastic electronic behaviors at extremely low temperature.

Table.I. Property of Binary pnictide semimetals XPm2

| Compound | Space group | Lattice Parameters | Property | Reference |
|---|---|---|---|---|
| NbAs$_2$ | $C12/m1$ | $a$=9.368Å<br>$b$=3.396Å<br>$c$=7.799Å | Topological semimetal;<br>Non-saturating large MR (2000%, 9T);<br>Negative longitudinal MR. | [20, 30] |
| NbSb$_2$ | $C12/m1$ | $a$=10.239(1)Å<br>$b$=3.630(1)Å<br>$c$=8.3285(2)Å | Topological semimetal;<br>Anisotropic giant MR (1200%, 9T);<br>High mobilities. | [31] |
| TaP$_2$ | $C12/m1$ | $a$=6.88(16)Å<br>$b$=3.2688(5)Å<br>$c$=7.4992(14)Å | Weak topological insulator;<br>Negative longitudinal MR. | [11] |
| TaAs$_2$ | $C12/m1$ | $a$=9.370Å<br>$b$=3.394Å<br>$c$=7.771Å | Topological semimetal;<br>Negative longitudinal MR ($10^6$%,9T);<br>High mobilities. | [14, 15, 18] |
| TaSb$_2$ | $C12/m1$ | $a$=10.222(4)Å<br>$b$=3.645(2)Å<br>c=8.292(3)Å | Anisotropic Magnetoresistivity;<br>Ultrahigh carrier mobility;<br>Negative longitudinal MR. | [10, 19] |
| VAs$_2$ | $C12/m$1 | $a$=9.05Å<br>$b$=3.27Å<br>$c$=7.46Å | Nodal-Line Semimetal;<br>Kondo Effect. | [17] |
| VSb$_2$ | $I4/mcm$ | $a$=6.5538Å<br>$b$=6.5538Å<br>$c$=5.6366Å | Large Magnetoresistance(447%,12T);<br>WAL. | [26] and our work. |